\documentclass[aps,preprint,showpacs]{revtex4}

\begin{document}

\title{Stable spatiotemporal solitons in Bessel optical lattices}
\author{D. Mihalache$^{1,2,3}$, D. Mazilu$^{1,3}$, F. Lederer$^{1}$, B. A.
Malomed$^{4}$, Y. V. Kartashov$^{2}$, L.-C. Crasovan$^{2,3}$, and
L. Torner$^{2}$} \affiliation{$^{1}$ Institute of Solid State
Theory and Theoretical Optics, Friedrich-Schiller Universit{\"a}t
Jena, Max-Wien-Platz 1, D-077743 Jena,
Germany\\
$^{2}$ICFO-Institut de Ciencies Fotoniques, and Department of Signal Theory
and Communications, Universitat Politecnica de Catalunya, 08034 Barcelona,
Spain\\
$^{3}$Department of Theoretical Physics, Institute of Atomic Physics, P.O.
Box MG-6, Bucharest, Romania\\
$^4$Department of Interdisciplinary Studies, Faculty of Engineering, Tel
Aviv University, Tel Aviv 69978, Israel}

\begin{abstract}
We investigate the existence and stability of three-dimensional
(3D) solitons supported by cylindrical Bessel lattices (BLs) in
self-focusing media. If the lattice strength exceeds a threshold
value, we show numerically, and using the variational
approximation, that the solitons are stable within one or
\emph{two} intervals of values of their norm. In the latter case,
the Hamiltonian-vs.-norm diagram has a ``swallowtail" shape, with
three cuspidal points. The model applies to Bose-Einstein
condensates (BECs) and to optical media with saturable
nonlinearity, suggesting new ways of making stable 3D BEC solitons
and ``light bullets" of an arbitrary size.
\end{abstract}

\maketitle

Nonlinear wave dynamics in lattice potentials has well-known realizations in
a wide range of physical settings, such as localized modes in solid-state
lattices \cite{Su} and waves in Bose-Einstein condensates (BECs) trapped in
optical lattices (OLs) \cite{BEC1}. Important realizations were predicted in
optics, in terms of the light transmission in arrays of coupled nonlinear
waveguides \cite{CJ} and optically-induced structures in photorefractive
(PR) materials (see surveys of recent results in Refs. \cite{Lederer} and
\cite{L4}), in layered slab media \cite{layered}, etc.
%and in multilayer fibers \cite{Fink}.
Stable solitary pulses trapped in lattices in dissipative media were also
predicted \cite{Staliunas}. Experiments demonstrate that lattices in PR
crystals indeed give rise to localized quasi-discrete dynamical patterns,
such as fundamental spatial solitons \cite{L1}, vortices \cite{L2}, and
soliton trains \cite{L3}.

Optical spatiotemporal solitons (alias ``light bullets")
%%\cite{Yuri}-\cite{TLB}
have attracted of lot of attention in the last several years, see
a review \cite{we}. They are 3D wave packets in which the
dispersion and diffraction are simultaneously balanced by
nonlinearity. The search for suitable media for the creation of
stable ``bullets" is a challenging problem, as multidimensional
solitons in Kerr-type focusing media are unstable against collapse
(blowup) \cite{Berge}. The same problem impedes creation of
multidimensional solitons in self-attractive BECs. Various schemes
to stabilize solitons in cubic media were proposed. In terms of
BECs, a promising avenue is the use of periodic OL potentials. 2D
and 3D solitons can be stabilized by OLs having the same dimension
\cite{L5}, or by \textit{low-dimensional} OLs, i.e., 2D and 3D
solitons may be stable in the presence of a 1D and 2D lattice,
respectively \cite{low-dim,L6}; a 1D lattice can support stable 3D
solitons in combination with the time-periodic alteration of the
nonlinearity sign, provided by the Feshbach resonance
\cite{Michal}. An essential advantage offered by the
low-dimensional OLs is that the solitons can freely move in the
unconfined direction(s), which opens a way to study their
collisions and bound states \cite{low-dim,Michal}. Essentially new
possibilities are opened in optics and BECs by radial OLs with
axial symmetry, that can be induced by diffraction-free
cylindrical Bessel beams \cite{Durnin}. Several 2D soliton
families, including fundamental solitons trapped at the center or
in different rings of the Bessel lattice (BL) and dipole-mode
solitons, were predicted in a cubic nonlinear medium, and were
shown to be stable in certain parameter domains \cite{Bessel}.
Very recently, stable ring-shaped solitons in cylindrical lattices
were also predicted in media with the saturable nonlinearity of
the PR type \cite{PLA}.

The difference between the previously studied 2D periodic OLs and the BL is
more than just another symmetry. First, the BL makes it possible to create a
soliton at a prescribed location (if the experimental field is large enough,
several solitons can be created and manipulated simultaneously). Moreover, each soliton can
be moved across the field by a slowly sliding laser beam.
Soliton networks and wires can be also created with arrays of Bessel beams,
a goal hard to achieve with harmonic lattices.
Quite
important is also the difference in the spatial scales. Indeed, the period
of OLs is dictated by the laser-beam's wavelength, $\lambda _{0}$
(although the period can be made larger by launching the interfering
waves under an angle to each other \cite{L4}). On the contrary, the radial
scale of the BL is completely independent of $\lambda _{0}$, being
determined by the phase mask through which the beam is passed. 
The use of the BL with an
essentially large scale will make it possible to create big \textit{paraxial
solitons} in optical media, and big solitons in BEC. The scale difference
also highlights the distinction of the BL from the single-mode optical
fibers, or cigar-shaped traps in BEC \cite{cigar}, where the radius is also
limited to microns.

The subject of this work are 3D solitons trapped in the BL, in the case of
the cubic self-attraction. This configuration can be directly implemented in
BECs with a negative scattering length, illuminated by an optical Bessel
beam. The corresponding normalized 3D Gross-Pitaevskii equation (GPE) for
the wave function $q$ is \cite{BEC1}
\begin{equation}
i\frac{\partial q}{\partial \xi }=-\frac{1}{2}\left(
\frac{\partial ^{2}q}{\partial {\eta }^{2}}+\frac{\partial
^{2}q}{\partial {\zeta }^{2}}+\frac{\partial ^{2}q}{\partial {\tau
}^{2}}\right) -|q|^{2}q-\Pi (r)q, \label{evolution}
\end{equation}where $\xi $ is time, $\tau $ and $\left( \eta ,\zeta \right) $ are,
respectively, the coordinates along the beam and in the transverse
plane, $r\equiv (\eta ^{2}+\zeta ^{2})^{1/2}$, and the potential
function $\Pi (r)$ is proportional to the local intensity of the
red-detuned optical wave which traps atoms at its anti-nodes
\cite{BEC1} [being interested in self-supporting solitons, we do
not include a magnetic (parabolic) trapping potential]. In this
work we mainly concentrate on the case of $\Pi (r)=pJ_{0}\left(
\sqrt{2\beta }r\right) $, with the Bessel function $J_{0}$
generated by the diffraction-free cylindrical beam \cite{Durnin}.
Constants $p$ and $\beta $ determine the strength and radial scale
of the BL. Note that the effective potential created by the Bessel
beam proper is proportional to $-J_{0}^{2}$, while the case of
$\Pi \sim J_{0}$ actually implies interference between the
cylindrical beam and a stronger plane wave with an amplitude
$A_{0}$ at the same frequency. Indeed, the latter configuration
implies the full field amplitude $A(r)=A_{0}+A_{1}e^{i\chi
}J_{0}\left( \sqrt{2\beta }r\right) $ (with constant $A_{1}$ and a
phase shift $\chi $), which generates the above potential through
the expansion $|A(r)|^{2}\approx A_{0}^{2}+2A_{0}A_{1}(\cos \chi
)J_{0}\left( \sqrt{2\beta }r\right) $. We demonstrate below that
the 3D solitons supported by the potentials $-J_{0}^{2}$ and
$-J_{0}$ may differ because of different degrees of localization
of these potentials. The latter potential (which was not proposed
before) may be more convenient, as it may be controlled by the
phase shift $\chi $; in particular, it is possible to flip the
sign of the $J_{0}$ potential, which is impossible for its
$J_{0}^{2}$ counterpart.
%Besides that, we will also
%demonstrate that the $J_{0}$ potential gives rise to an
%essentially larger stability region for 3D solitons than its
%$J_{0}^{2}$ counterpart.

In nonlinear optics, $\xi $, $\left( \eta ,\zeta \right) $, and
$\tau $ are, respectively, normalized propagation and transverse
coordinates and local time \cite{we}, assuming anomalous chromatic
dispersion. Then, Eq. (\ref{evolution}) governs the nonlinear
transmission of a probe signal in a strong BL induced by a pump
beam launched in an orthogonal polarization, or at a different
wavelength. In this case, $\Pi (r)$ measures local modulation of
the refractive index for the probe signal induced by the pump beam
through the cross-phase modulation, and the choice of $\Pi \sim
J_{0}$ again implies the interference between the Bessel beam and
a stronger plane wave. The pump beam is assumed to propagate in a
nearly linear regime, that actually implies strong saturation of
the pump beam in the medium with saturable nonlinearity.

Equation (\ref{evolution}) conserves the norm, $U~=~\int \int \int
\left\vert q(\eta ,\zeta ,\tau )\right\vert ^{2}d\eta d\zeta d\tau
$, Hamiltonian, $H$, and angular momentum. We search for
stationary solitons, $q(\eta ,\zeta ,\tau ,\xi )=w(r,\tau )\exp
(ib\xi )$, where $b$ is a real propagation constant, and $w$,
which must vanish at $|\tau |,r\rightarrow \infty $, obeys the
equation
\begin{equation}
w_{\tau \tau }+w_{rr}+r^{-1}w_{r}+2[p\Pi (r)-b]w+2w^{3}=0.
\label{stationary}
\end{equation}In numerical calculations, we fix the radial scale, $\beta \equiv 10$, and
vary $b$ and $p$ (cf. Ref. \cite{Bessel}).

Using an obvious Lagrangian representation of Eq. (\ref{stationary}),
%%\begin{equation}
%%L=2\pi \int_{0}^{\infty }rdr\int_{-\infty }^{+\infty }
%%d\tau \left[ \frac{1}{2%%}\left( w_{\tau }^{2}+w_{r}^{2}-w^{4}\right)
%%+\left( b-p\Pi (r)\right) w^{2}%%\right] .  \label{L}
%%\end{equation}
we start with the variational approximation (VA). To this end, we adopt the
Gaussian \textit{ansatz} for the 3D soliton, $w(r,\tau )=A\exp \left[
-\left( \tau ^{2}/T^{2}+r^{2}/R^{2}\right) /2\right] $, with free parameters
$T$, $R$ and $A$. The known procedure \cite{VA} leads to
%%The substitution of this in Eq. (\ref{L}) makes it possible to
%%calculate the Lagrangian, which coincides with the value of the
%%Hamiltonian $H$ for the above \textit{ansatz} :
%%\[
%%L=H=\frac{U}{2}\left[ b-p\exp \left( -\frac{\beta R^{2}}{2}\right)
%%+\frac{1}{%%4T^{2}}+\frac{1}{2R^{2}}-\frac{U}{4\sqrt{2}
%%\pi ^{3/2}TR^{2}}\right] ,\]%%The variational
%%equations, $\partial L/\partial \left( U,T,R\right) =0$, lead to the
relations $T^{2}=R^{2}/\left( 1-p\beta R^{4}e^{-\beta R^{2}/2}
\right) $, $b=p\left( 1-3\beta R^{2}/4\right) e^{-\beta R^{2}/2}+
\left(4R^{2}\right)^{-1}$, and $U^{2}=8\pi ^{3}R^{2}\left( 1-
p\beta R^{4}e^{-\beta R^{2}/2}\right)$, where $U=\pi
^{3/2}A^{2}TR^{2}$ is the soliton's norm. These equations predict
the main characteristic of the soliton family, $U(b)$. Then, the
Vakhitov-Kolokolov (VK) criterion, $dU/db>0 $, provides for
stability of the solitons against perturbations with real
eigenvalues \cite{Berge}. A first VK-stable interval appears
around $R=R_{0}^{(1)}\equiv 2.75/\sqrt{\beta }$, when the BL
strength attains a threshold value, $p_{\mathrm{thr}}^{(1)}\approx
0.12\beta $. A \emph{second stability interval} appears around
$R_{0}^{(2)}\equiv \left[ 2\left( 3-\sqrt{3}\right) /\beta \right]
^{1/2}\approx 1.\,\allowbreak 59/\sqrt{\beta }$, when $p$ passes a
\emph{second threshold}, $p_{\mathrm{thr}}^{(2)}\approx 0.32\beta
$. As shown below, numerical results confirm the existence of two
stability intervals for $p>p_{\mathrm{thr}}^{(2)}$. It is relevant
to compare the above values $R_{0}^{(1,2)}$ and the size of the
central core of the BL potential in Eq. (\ref{stationary}), which
is determined by the first zero of the function $J_{0}\left(
\sqrt{2\beta }r\right) $, $r_{0}\approx \allowbreak
1.\,\allowbreak 70/\sqrt{\beta }$. The comparison suggests that
the above simple ansatz, that neglects BL-induced fringes in the
shape of the soliton, is appropriate, as the predicted solitons do
not cover many rings of the radial lattice.

Numerically, soliton families were found from Eqs.
(\ref{stationary}) by means of the relaxation method. Simulations
of the full GPE (\ref{evolution}) were carried out using the
Crank-Nicholson scheme. To test the stability, we took perturbed
solutions as $q=[w(r,\tau )+u(r,\tau )\exp (\delta \xi +in\phi
)+v^{\ast }(r,\tau )\exp (\delta ^{\ast }\xi -in\phi )]\exp (ib\xi
)$, where $\phi $ is the azimuthal angle in the plane $\left( \eta
,\zeta \right) $, and $n=0,1,2,...$ . The linearization of Eq.
(\ref{evolution}) leads to straightforward equations for the
perturbation mode $(u,v)$ and its
growth rate $\delta $. %%\begin{eqnarray}
%%&&i\delta u+\frac{1}{2}\left( u_{\tau \tau
%%}+u_{rr}+r^{-1}u_{r}-n^{2}r^{-2}u\right)   \nonumber \\
%%+\left[ p\Pi (r)-b\right] u+w^{2}(v+2u) &=&0,  \nonumber \\
%%&&-i\delta v+\frac{1}{2}\left( v_{\tau \tau
%%}+v_{rr}+r^{-1}v_{r}-n^{2}r^{-2}v\right)   \nonumber \\
%%+\left[ p\Pi (r)-b\right] v+w^{2}(u+2v) &=&0,  \label{growth}
%%\end{eqnarray}that were solved numerically.
The numerical results are summarized in Fig. 1. We found that the
3D solitons exist in the $J_{0}$ potential if the propagation
constant $b$ exceeds a certain cutoff value $b_{\mathrm{co}}$.
Figure 1(a) shows that $b_{\mathrm{co}}$ increases with the BL
strength $p$ [$b_{\mathrm{co}}(p)$ vanishes at $p=0$ as unstable
free-space 3D solitons exist for all $b>0$]. The soliton's norm
$U$ is a \emph{non-monotonous} function of $b$, see Fig. 1(b). The
prediction of the VK criterion, that parts of the soliton family
with $dU/db>0$ may be stable, was verified by numerical
computation of the growth rates, $\mathrm{Re}(\delta )$. It was
found that the solitons are completely stable \emph{precisely} in
regions singled out by the condition $dU/db>0$, see Fig. 1(c). For
the $J_{0}$ potential and relatively small values of the BL
strength $p$, the solitons are stable in a narrow interval of $b$
to the right of $b_{\mathrm{co}}$ [see a small solid segment of
the curve in Fig. 1(b) corresponding to $p=5$]. Losing their
stability for $b$ exceeding the value at which $dU/db$ vanishes,
the 3D solitons drastically differ from their BL-supported 2D
counterparts, that may be stable in their entire existence domain
if the lattice strength $p$ is large enough \cite{Bessel}.
Remarkably, for moderate values of $p$, the 3D solitons are stable
in \emph{two disjoint intervals} of $b$ [one abuts on the cutoff
point, $b=b_{\mathrm{co}}$, and the other is found at relatively
large values of $b$, see the curve corresponding to $p=5.5$ in
Fig. 1(b)]. This feature was predicted above by the VA. However,
there is \emph{only one} soliton-stability domain, adjacent to the
cutoff, for the $J_{0}^{2}$ potential, see Fig. 1(d). The
difference can be traced to different localization degrees of the
$J_{0}^{2}$ and $J_{0}$ potentials.

The Hamiltonian-vs.-norm diagrams for the soliton families, which is a
useful tool for the analysis of the soliton stability \cite{Kusmartsev}, are
plotted in Fig. 2. With the $J_{0}$ potential, they exhibit one or three
cuspidal points. In the latter case [Fig. 2(a)], the diagram features a
``swallowtail" pattern, which accounts for the existence of the two
above-mentioned distinct stability regions for the 3D solitons. Although
this pattern is one of generic possibilities known in the catastrophe
theory, it rarely occurs in physical models \cite{Kusmartsev}.
%%A review of applications of the catastrophe theory to the
%%soliton-stability problem, based on the Whitney theorem
%%\cite{Whitney} for two-dimensional maps, which are constructed with
%%the aid of the dynamical
%%invariants of the nonlinear evolution equations, can be found in Ref.
%%\cite{Kusmartsev}.
In Fig. 3 we display three typical examples of stable 3D solitons
supported by the $J_{0}$ potential. A low-amplitude broad soliton
covering several lattice rings, with the propagation constant $b$
close to the cutoff $b_{\mathrm{co}}$, is shown in Fig. 3(a), and
high-amplitude narrow solitons, mostly trapped within the BL core,
are presented in Figs. 3(b,c).

An important issue for the 3D solitons in the focusing medium is the
occurrence of collapse. We expect that the solitons which are unstable
against small perturbations either undergo the collapse, or decay into
radiation waves, depending on the perturbation. To test these expectations,
we simulated the evolution of perturbed solitons, starting with $q(\xi
=0)=w(\eta ,\zeta ,\tau )(1+\epsilon f)$, where $\epsilon $ is a small
perturbation amplitude, and $f$ is either a random variable uniformly
distributed in the interval $[-1,1]$, or constant, $f=1$. We have found that
the solitons which were predicted to be stable in terms of the perturbation
eigenvalues are indeed stable in the direct simulations. As an example, in
Fig. 4 we show the evolution of a soliton with a random perturbation whose
amplitude is $\epsilon =0.05$. Despite a considerable transient change of
its shape, the soliton relaxes to a self-cleaned form which is quite close
to the unperturbed one. Under the same random perturbation, linearly
unstable solitons were found to decay into radiation. The uniform initial
perturbation, $f=1$, amounts to an instantaneous change of the soliton's
amplitude. This perturbation excites persistent oscillations of linearly
stable solitons, while the effect on linearly unstable ones is completely
different from the response to random perturbations: they do not decay, but,
instead, blow up.

In conclusion, we have constructed families of 3D solitons supported by zero-th order 
Bessel optical lattices. We have found that, in one or \emph{two}
intervals, the solitons are stable, provided that the lattice is
strong enough. 
We also expect that 3D solitons can be stabilized by higher-order Bessel lattices, too.
This is the first example of a 3D soliton
stabilized by an axially symmetric lattice; unlike stable 3D
solitons in 2D periodic lattices, the soliton's size is not
determined by the wavelength of the laser beam, but may be chosen
at will. Depending on the lattice strength, the
Hamiltonian-vs.-norm diagram for solitons in the $J_{0}$ potential
displays a ``swallowtail" bifurcation, a rare phenomenon in
physics.

The results suggest new approaches to the still unresolved \cite{we} problem
of the creation of 3D solitons in the experiment. One possibility is to make
a BEC soliton in the cylindrical optical lattice, which may be controlled by
the superposition of the Bessel beam and uniform background field. In
optics, creation of the 3D lattice solitons require the bimodal setting in a
medium with saturable nonlinearity. In all these settings, stable 3D
solitons are predicted to feature universal properties.

Support from Deutsche Forschungsgemeinschaft (DFG), Instituci{\'{o} }
Catalana de Recerca i Estudis Avan{\c{c}}ats (ICREA), and the Israel Science
Foundation (grant No. 8006/03) is acknowledged.

\newpage

\begin{figure}[tbp]
%\center{\includegraphics[width=4in]{figure1.eps}}
\caption{(a) The propagation constant cutoff vs. the lattice strength, shown
for the cylindrical potentials of both types, $J_{0}$ and $J_{0}^{2}$. (b,d)
The soliton's norm vs. the propagation constant for the potentials $J_{0}$
and $J_{0}^{2}$, respectively. (c) The perturbation growth rate vs. the
propagation constant for the $J_{0}$ potential. Here and below, solid and
dashed curves correspond, respectively, to stable and unstable branches.}
\label{Fig.1}
\end{figure}

\begin{figure}[tbp]
%\center{\includegraphics[width=4in]{figure2.eps}}
\caption{The Hamiltonian-vs.-norm diagrams for $p=5.5$ (a) and $8$ (b). Here
and below, all the figures are shown for the $J_{0}$ potential.}
\label{Fig.2}
\end{figure}

\begin{figure}[tbp]
%\center{\includegraphics[width=4in]{figure3.eps}}
\caption{Isosurface plots for stable solitons: (a) $p=5$,
$b=0.11$, $U=6.477$; (b) $p=5.5$, $b=0.85$, $U=3.677$; (c) $p=6$,
$b=2$, $U=3.591$.} \label{Fig.3}
\end{figure}

\begin{figure}[tbp]
%\center{\includegraphics[width=4in]{figure4.eps}}
\caption{The evolution of a stable soliton under a random
perturbation, for $p=5$, $b=0.11$, $U=6.477$: (a) $\protect\xi
=0$, (b) $\protect\xi =200$, (c) $\protect\xi =500$. }
\label{Fig.4}
\end{figure}

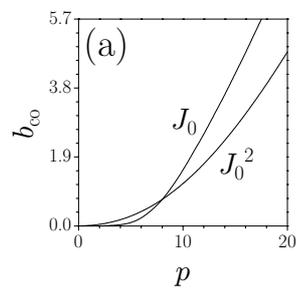
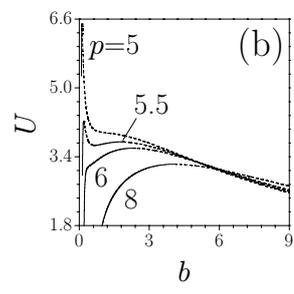
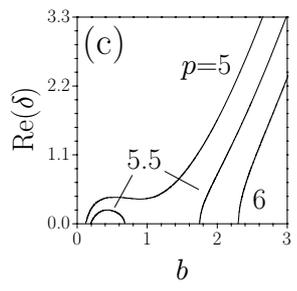
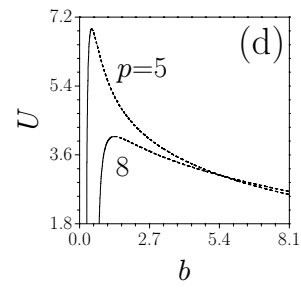
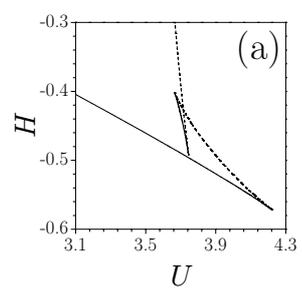
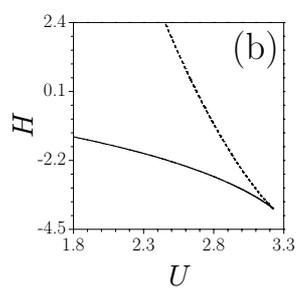
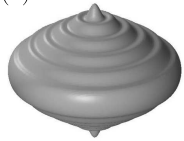
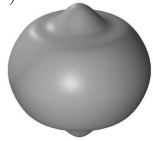
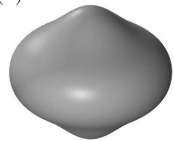
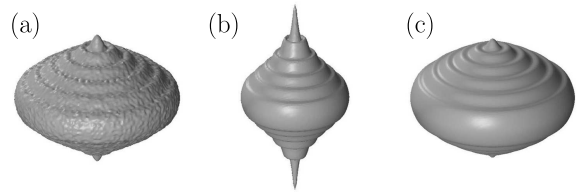
\end{document}